\documentclass{acmart}
\AtBeginDocument{%
  }

\setcopyright{acmlicensed}
\copyrightyear{2026}
\acmYear{2026}
\acmDOI{XXXXXXX.XXXXXXX}
\acmConference[Conference acronym 'XX]{Make sure to enter the correct
  conference title from your rights confirmation email}{June 03--05,
  2018}{Woodstock, NY}
\acmISBN{978-1-4503-XXXX-X/2018/06}

\usepackage{algpseudocode}
\usepackage{algorithm}
\usepackage{braket}
\usepackage{xspace}
\usepackage{subcaption}

\newcommand{\ketb}[1]{\ensuremath{\ket{\mathbf{b}_{#1}}}\xspace}
\newcommand{\brab}[1]{\ensuremath{\bra{\mathbf{b}_{#1}}}\xspace}

\begin{document}

\title{A scalable sample-based quantum diagonalization solver for general Pauli-sum Hamiltonians}

\author{Yutaro Iiyama}
\correspondingauthor
\email{iiyama@icepp.s.u-tokyo.ac.jp}
\orcid{0000-0002-8297-5930}
\affiliation{%
  \institution{International Center for Elementary Particle Physics, The University of Tokyo}
  \city{Bunkyo}
  \state{Tokyo}
  \country{Japan}
}








\renewcommand{\shortauthors}{Iiyama}

\begin{abstract}
A scalable ground-state solver to be used in sample-based quantum diagonalization (SQD) is presented. The solver is built on JAX API, and is intended to be run on GPU platforms and possibly many-GPU high-performance computing facilities. It is capable of subspace diagonalization in up to $2^{30}$-dimensional subspace, with possible upgrades to enable even larger-scale operations. The two main components, a general-purpose eigensolver and an SQD-specific matrix-vector operation function, are described in detail. Efficiency of the eigensolver with respect to the GPU memory is demonstrated using a simple physics model. Furthermore, good strong scaling of the eigensolver in many-GPU environment is observed.
\end{abstract}




\maketitle

\section{Introduction}

Sample-based quantum diagonalization (SQD) has emerged as one of the most promising application of near-term quantum computers\cite{Kanno2026-sd,Robledo-Moreno2025-pi,Sugisaki2025-xp,Yu2025-gw,Shirakawa2025-no}. Regarded as a superior alternative to the variational quantum eigensolver\cite{Peruzzo2014-nx}, SQD has been applied to a wide variety of ground-state problems of quantum many-body systems, demonstrating energy approximation accuracies comparable to the best classical methods.

In SQD, a many-body Hamiltonian of an intractable dimensionality is projected onto a small subspace, in which the ground-state eigenpair of the reduced Hamiltonian is found. The projection subspace dimensions are identified through sampling quantum states realized on a quantum computater. In the original formulation, which is also known as quantum selected configuration interaction (QSCI), these states are the approximations of the ground state prepared with ansatz circuits. In the variant dubbed sample-based Krylov quantum diagonalization (SKQD), final states of time evolutions under the target Hamiltonian for multiple durations are sampled instead. Crucially, in all versions of SQD, the only role quantum computers play is to suggest the subspace dimensions. Everything else, including correcting potential errors in the obtained samples, is performed as classical computation. This separation of responsibility between quantum and classical computation, with a strong emphasis on the latter, is the innovation that allows SQD to be a candidate framework in which quantum advantage is achieved even with pre-fault-tolerant quantum computers.

It is of critical importance that ``small'' is a relative term in this context; in practice, the number of subspace dimensions can reach up to tens of billions. Even when the number of measurement outcomes from the quantum computer are less than a million, classical postprocessing can drastically enlarge the subspace size by e.g. identifying additional relevant dimensions through symmetry considerations. The implication is that the Hamiltonian projection and diagonalization is not performed sequentially, because it is impossible to store the reduced Hamiltonian matrix in memory unless it is known to be extremely sparse. Instead, a matrix-free method must be employed for diagonalization, in which the matrix-vector operation, which computes the result of multiplying a vector with the reduced Hamiltonian, must perform the projection in situ. Additionally, this computation would likely be executed in distributed parallel environments such as high-performance computing (HPC) facilities.

The central software components for large-scale SQD are therefore a matrix-free eigensolver and a specialized matrix-vector operation. Since both happen to involve many vector-vector and matrix-vector multiplications, they are amenable to GPU acceleration. In fact, as modern HPC architectures provide the majority of their compute through GPUs, it is essentially necessary that these components are implemented in a GPU-native manner.

There already exist at least two such implementations that are open source. The first is SBD-G\cite{Doi2026-vs}, which is a header-only C++ library utilizing Thrust\cite{thrust} and MPI to enable multi-GPU parallelism. This library specializes in SQD of chemistry and hard-core boson Hamiltonians, and has achieved state-of-the-art performance of diagonalizing a $1.36 \times 10^{18}$-dimensional Hamiltonian projected onto a $2.34 \times 10^{10}$-dimensional subspace using 128 GPU nodes in Supercomputer Miyabi\cite{Hanawa2024-gd}.

The second implementation, which is more similar in scope and design to what is presented in this paper, is in Ref.~\cite{cudaq} as a part of the demonstration of the functionalities of the CUDA Quantum library. Unlike SBD-G, the Hamiltonian it works with is expressed as a general Pauli sum of up to 64 qubits. The eigensolver is based on the thick-restarted Lanczos method, which requires retention of many large vectors in memory.

Our SQD solver, which is also open source, is written in Python and built on the JAX API\cite{jax2018github}. Like the CUDA Quantum implementation, it targets general Pauli-sum Hamiltonians, but there is no limit to the number of qubits the Hamiltonians act on. The eigensolver employs the locally optimal conjugate gradient method, and is optimized for GPU memory footprint. By taking advantage of the built-in parallelization capabilities in JAX, the full solver can be run in many-GPU HPC as well as a single workstation, with minimal change in the user code.

We designed the library to make the eigensolver a standalone general utility which can be used to find the lowest few eigenpairs of any Hermitian operator, as long as its operation on a vector is implemented as a JAX just-in-time compiled function. This eigensolver can be scaled up to eigenvalue problems of $10^{10}$ or higher dimensions. The SQD solver is a downstream user of this eigensolver, providing the SQD-specific matrix-vector operator constructed from the problem Hamiltonian and the subspace dimensions. This matrix-vector operator currently throttles the scaling of the solver, as it has a hard limit of $2^{31}$ for the number of subspace dimensions that can be handled. An upgrade to remove this restriction is planned.

This paper presents our SQD solver with the goal of broadening its user base by clarifying its internal workings and demonstrating its capabilities. We describe the conjugate gradient eigensolver and matrix-vector operation in Section~\ref{sec:components}. The performance and scaling of the standalone eigensolver and the overall SQD solver are discussed in Section~\ref{sec:demonstration}. Section~\ref{sec:conclusion} concludes with comments on the planned extensions.

\section{SQD components}\label{sec:components}

\subsection{Eigensolver}\label{subsec:eigensolver}

The locally optimal conjugate gradient method for an eigenvalue problem
\begin{equation}\label{eq:eigenvalue_problem}
    A v = \lambda_0 v
\end{equation}
 computes a series of solution estimates $\{v^{(0)}, v^{(1)}, \dots\}$ iteratively, with progressively better approximations. Here, $A$ is an $D \times D$ symmetric or Hermitian matrix and $\lambda_0$ is its lowest eigenvalue. At each iteration, the new estimate is obtained as the minimizer of the Rayleigh quotient of $A$ within the three-dimensional subspace spanned by the current and previous estimates and the current residual. Derivation and detailed discussion of the algorithm, which has been known for several decades, can be found elsewhere\cite{Albrecht2013-ti}. We also provide our implementation as a pseudocode in Appendix~\ref{app:locg}.

This algorithm can be regarded as a single-vector unpreconditioned version of the better-known locally optimal block preconditioned conjugate gradient (LOBPCG) method, which can identify multiple lowest or highest eigenpairs of $A$ simultaneously. We chose to create our own eigensolver, even though there exist multiple public implementations of LOBPCG, in order to minimize the GPU memory footprint. By focusing on the lowest eigenpair only and aggressively reusing memory, our eigensolver can complete its calculation with as low as five $D$-dimensional vectors simultaneously materialized in the GPU RAM. This means that Equation~\eqref{eq:eigenvalue_problem} with $D=10^{9}$ and $v$ represented as an array of 128-bit floating point complex numbers may be solved on a single GPU with a $100\,\mathrm{GB}$ memory. Note however that the actual maximum number of vectors in RAM can be larger, depending on the matrix-vector operation function that encodes $v \mapsto Av$.

When tackling a higher-dimensional problem or using a matrix-vector function that requires a large RAM allocation, the eigensolver can be executed in parallel computing environments by splitting the $D$-dimensional vectors into chunks that are distributed among the parallel nodes. There is thus in principle no limit to the problem dimension the eigensolver can handle, although in reality the maximum number of parallel GPU nodes available at HPC centers are typically at a few hundreds, implying a dimensionality limit of about $10^{10}$. For a fixed problem size, the eigensolver is expected to show a good strong scaling, as the only operations it performs on the vectors outside the matrix-vector function are element-wise additions, scalar multiplications, and vector inner products, all of which entail minimal data transfer between the parallel nodes.

It should be mentioned that we do also provide an option to compute the second and higher eigenpairs by passing to the eigensolver a list of vectors that the solution should be orthogonalized against. This crude approach introduces minimal additional memory footprint but is known to suffer from numerical instability, which is exactly why LOBPCG was invented. For SQD, this functionality is not used at any rate.

\subsection{Reduced Hamiltonian matrix-vector operation}
\subsubsection{Algorithm}\label{subsubsec:matvec_algorithm}

We now discuss the SQD-specific matrix-vector operation function. First, we must clarify what exactly correspond to the $A$ and $v$ of Equation~\eqref{eq:eigenvalue_problem} in the case of SQD. Let $B = \{\mathbf{b}_{n}\}_{n=1}^{N}$ be the $q$-bit binary indices (bit strings) of the subspace dimensions identified by quantum states sampling and postprocessing. Using the projector onto this subspace
\begin{equation}
    P := \sum_{n=1}^{N} \ketb{n} \brab{n},
\end{equation}
the subspace diagonalization of the $q$-qubit Hamiltonian $H$ is formulated as the process of identifying $\alpha := (\alpha_n)_{n=1}^{N}$ and $\tilde{E}_0$ in
\begin{equation}
    P H P \sum_{n=1}^{N} \alpha_n \ketb{n} = \tilde{E}_0 \sum_{n=1}^{N} \alpha_n \ketb{n}.
\end{equation}
The eigenvalue problem to solve is therefore
\begin{equation}
    \sum_{n=1}^{N} \brab{m} H \ketb{n} \alpha_{n} = \tilde{E}_0 \alpha_{m},
\end{equation}
which means that the matrix-vector operation must compute
\begin{equation}\label{eq:sqd_matvec}
    \beta := (\beta_{m})_{m=1}^{N} = \left( \sum_{n=1}^{N} \brab{m} H \ketb{n} \alpha_{n} \right)_{m=1}^{N}.
\end{equation}

The Hamiltonian $H$ is a Pauli sum
\begin{equation}\label{eq:hamiltonian_breakdown}
    H = \sum_{k=1}^{K} c_k P_k,
\end{equation}
where each Pauli string $P_k \in \{I, X, Y, Z\}^{\otimes q}$ is a tensor product of $q$ Pauli operators. We break down Equation~\eqref{eq:sqd_matvec} using \eqref{eq:hamiltonian_breakdown} and compute each $\brab{m} P_k \ketb{n}$ separately. To do so, $P_k$ is expressed in the symplectic representation
\begin{equation}\label{eq:symplectic_rep}
    P_k = (-i)^{\mathbf{x}_k \cdot \mathbf{z}_k} \bigotimes_{j=0}^{q-1} Z^{z_{k,j}} X^{x_{k,j}},
\end{equation}
where $\mathbf{x}_k = (x_{k,j})_{j=0}^{q-1}, \mathbf{z}_k = (z_{k,j})_{j=0}^{q-1} \in \{0,1\}^{q}$ are the X and Z signatures of $P_k$. The phase factor $(-i)^{\mathbf{x}_k \cdot \mathbf{z}_k}$ is static and can be combined with the Pauli coefficient into
\begin{equation}
    \tilde{c}_k := (-i)^{\mathbf{x}_k \cdot \mathbf{z}_k} c_k.
\end{equation}
Looking at the matrix elements of the operators, because $Z$ is diagonal in the computational basis and $X$ maps one basis state to another, we have
\begin{equation}\label{eq:matrix_element}
    \brab{m} \bigotimes_{j=0}^{q-1} Z^{z_{k,j}} X^{x_{k,j}} \ketb{n} = (-1)^{\mathbf{b}_m \cdot \mathbf{z}_k} \delta_{(\mathbf{b}_m \oplus \mathbf{x}_k), \mathbf{b}_n}.
\end{equation}
Therefore, the projection of the Hamiltonian boils down to identifying a $\mathbf{b}_{n_m}$ for each $\mathbf{b}_m$ such that
\begin{equation}\label{eq:xsource}
    \mathbf{b}_{n_{k,m}} = \mathbf{x}_k \oplus \mathbf{b}_m.
\end{equation}
There is at most one such index in $B$.

We use Algorithm~\ref{alg:xsource} (X source index search) to find $n_{k,m}$ for each $m$. This is an application of a search algorithm designed for fixed tensor size architectures (e.g. GPU). The returned vector of indices $(n_{k,m})_{m=1}^{N}$ may contain invalid entries, indicating that bit strings satisfying Equation~\eqref{eq:xsource} do not exist in $S$ for those slots.

\begin{algorithm}
\caption{X source index search. In the lines, $\text{StableSort}$ refers to a sorting that keeps the original relative order of elements that have equal values, and $\mathrm{arg}\text{StableSort}$ returns the indices into the original array which would result in the sorted array.}\label{alg:xsource}
\begin{algorithmic}[1]
\Require Sorted list of binary indices $S=(\mathbf{b}_m)_{m=1}^{N}$, binary X signature $\mathbf{x}$
\Ensure List of indices $(n_m)_{m=1}^{N}$, taking values in $\{0, \dots, N\}$, such that $\mathbf{b}_{n_m} = \mathbf{x} \oplus \mathbf{b}_{m}$ if $n_m \neq 0$.

\Statex \textit{Concatenate $S$ and $\mathbf{x} \oplus S$ and take the sorting indices}
\State $T \gets (\mathbf{x} \oplus \mathbf{b}_{m})_{m=1}^{N}$
\State $U \gets S \mathbin{\|} T$
\State $P \gets \Call{StableSort}{U}$
\State $Q \gets \mathrm{arg}\Call{StableSort}{U}$

\Statex \textit{Collect $(n_m)_{m=1}^{N}$ from $Q$ by checking repetitions in $P$}
\ForAll{$m \in \{1, \dots, N\}$}
  \State $k_m \gets k \in \{1, \dots, 2N\} \; \text{s.t.} \; Q[k] = m$
  \If{$P[k_m + 1] = P[k_m]$}
    \State $n_m \gets Q[k_m + 1] - N$
  \Else
    \State $n_m \gets 0$
  \EndIf
\EndFor
\State \Return $(n_m)_{m=1}^{N}$
\end{algorithmic}
\end{algorithm}

Plugging Equation~\eqref{eq:hamiltonian_breakdown}-\eqref{eq:matrix_element} back into Equation~\eqref{eq:sqd_matvec}, we get
\begin{equation}\label{eq:betam_as_ksum}
    \beta_m = \sum_{k=1}^{K} \beta_{k,m},
\end{equation}
where
\begin{equation}\label{eq:pkm}
    \beta_{k,m} = \begin{cases}
        0 & \text{if}\; n_{k,m} = 0 \\
        (-1)^{\mathbf{b}_m \cdot \mathbf{z}_k} \tilde{c}_k \alpha_{n_{k,m}} & \text{otherwise}
    \end{cases}.
\end{equation}
In practice, the ``take'' function is used on $\alpha$ for indices $n_{k,m}$ to obtain the right-hand side of Equation~\eqref{eq:pkm}, with a fill-in value of 0 for invalid indices.

Because X source index search and the ``take'' function are both expensive operations, the latter especially when the $\alpha$ vector is distributed, we seek to minimize their usage by grouping the Hamiltonian terms by the X signature. Equation~\eqref{eq:hamiltonian_breakdown} and \eqref{eq:symplectic_rep} combined is rewritten as
\begin{equation}
    H = \sum_{\ell=1}^{L} \left( \sum_{k \in G_{\ell}} \tilde{c}_{k} \bigotimes_{j=0}^{q-1} Z^{z_{k,j}} \right) \bigotimes_{j=0}^{q-1} X^{x_{(\ell),j}},
\end{equation}
where $\{\mathbf{x}_{(\ell)}\}_{s=1}^{L}$ are the $L$ distinct X signatures found in the Hamiltonian, and $G_{\ell}$ is the set of indices for the term in the original breakdown \eqref{eq:hamiltonian_breakdown} with the X signature $\mathbf{x}_{(\ell)}$. With this grouping, we compute the X source indices $\{n_{(\ell),m}\}_{m=1}^{N}$ for each $\mathbf{x}_{(\ell)}$. Equation~\eqref{eq:betam_as_ksum} is then rewritten as
\begin{equation}
    \beta_m = \sum_{\ell=1}^{L} \beta_{(\ell),m}
\end{equation}
with
\begin{equation}
    \beta_{(\ell),m} = \begin{cases}
        0 & \text{if}\; n_{(\ell),m} = 0 \\
        \left(\sum_{k \in G_{\ell}} (-1)^{\mathbf{b}_m \cdot \mathbf{z}_{k}} \tilde{c}_k \right) \alpha_{n_{(\ell),m}} & \text{otherwise}.
    \end{cases}
\end{equation}

\subsubsection{Caching static data}

The matrix-vector operation of Section~\ref{subsubsec:matvec_algorithm} is called hundreds of times during the eigensolver execution in some cases. Only the $\alpha$ vector changes between the invocations, while the diagonals
\begin{equation}
    \{d_{(\ell),m}\}_{\ell=1,m=1}^{L,N} := \left\{\left(\sum_{k \in G_{\ell}} (-1)^{\mathbf{b}_m \cdot \mathbf{z}_{k}} \tilde{c}_k \right)\right\}_{\ell=1,m=1}^{L,N}
\end{equation}
and the source indices $\{n_{(\ell),m}\}_{\ell=1,m=1}^{L,N}$ stay constant. Therefore, computing these quantities once and reusing them can speed up the diagonalization process, with a potential cost in GPU memory. We thus provide such caching options in our matrix-vector function.

If each $d_{(\ell),m}$ is represented with a 128-bit complex number, storing the diagonals in memory would require $16 LN$ bytes. For the source indices, the memory requirement would be $4 LN$ bytes, if 32-bit integers are used (which is actually necessary, as explained in the following). Depending on the number of distinct X signatures $L$, these costs can be prohibitive. Given that X source finding is significantly expensive compared to all other operations in the diagonalization process, the default setting is to cache the X source indices but compute the diagonals on the fly at each invocation of the matrix-vector function. Note, though, that if the diagonals are also cached, the list of dimension indices $S$ will no longer be necessary to be kept in memory, freeing up $\lceil \frac{q}{8} \rceil N$ bytes. Therefore, if $q > 128 L$, it is strictly advantageous to cache also the diagonals.

\subsubsection{Limitation on the number of subspace dimensions}

As evident from Algorithm~\ref{alg:xsource}, X source search is heavily reliant on stable sorting of subspace dimension indices. Since stable sorting of more than $2^{31} - 1$ elements is currently not implemented in JAX, while Algorithm~\ref{alg:xsource} involves sorting a concatenated array of $2N$ elements, the maximum number of subspace dimensions $N$ is currently $2^{30} - 1$.

However, global sorting of the binary indices is not a strict necessity. When the $\alpha$ vector is distributed among parallel compute nodes, we can foresee defining separate X source indices for each node, in which case Algorithm~\ref{alg:xsource} would be run on chunks of $S$ in parallel. Although using such indices can substantially slow down the function execution, we will be able to work around the current limitation this way.

\section{Performance of the SQD solver}\label{sec:demonstration}

\subsection{Eigensolver}

To validate the conjugate gradient eigensolver, we compute the two lowest energy eigenvalues $E_0, E_1$ of the one-dimensional transverse-field Ising model (TFIM) with $q=30$ sites and a periodic boundary condition. Explicitly, the Hamiltonian of the model is
\begin{equation}
    H = -\sum_{j=0}^{q-1} Z_{j} Z_{j+1} + h \sum_{j=0}^{q-1} X_j,
\end{equation}
where $Z_{q}$ is identified with $Z_{0}$ and $h$ is the transverse field parameter. This model is known to exhibit a phase transition at $h=1$ in the thermodynamic limit, in which the ground state is doubly degenerate in the ordered ($h < 1$) phase and is unique in the disordered ($h > 1$) phase. 

\begin{figure}
    \centering
    \begin{subfigure}{0.48\textwidth}
        \centering
        \includegraphics[width=\textwidth]{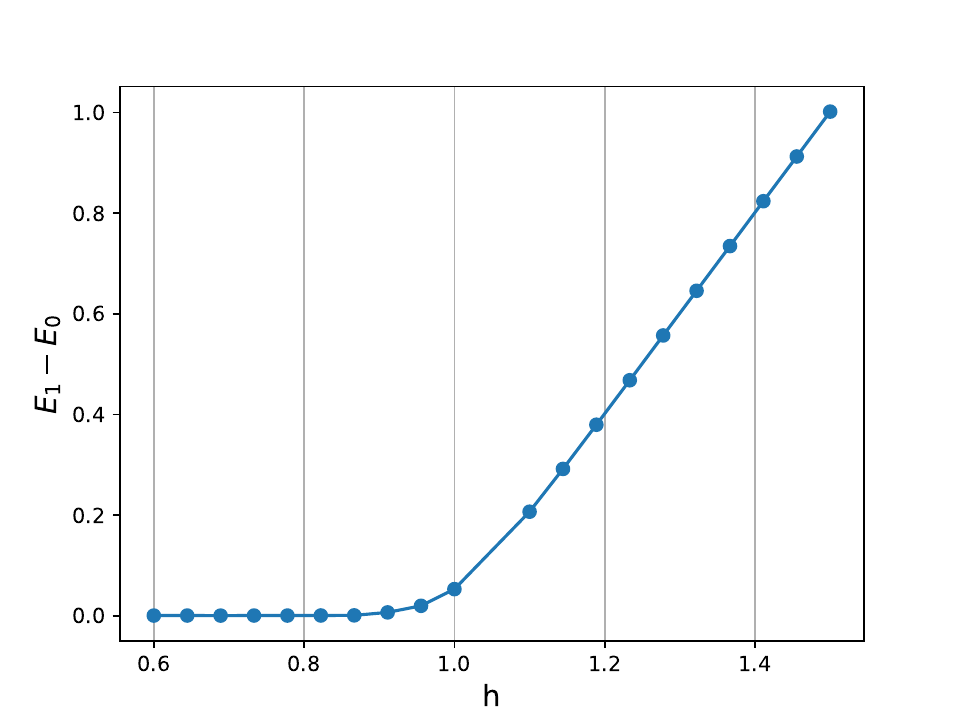}
        \caption{30-site TFIM energy gap as a function of $h$.}
        \label{fig:ising_gap}
    \end{subfigure}
    \hfill
    \begin{subfigure}{0.48\textwidth}
        \centering
        \includegraphics[width=\textwidth]{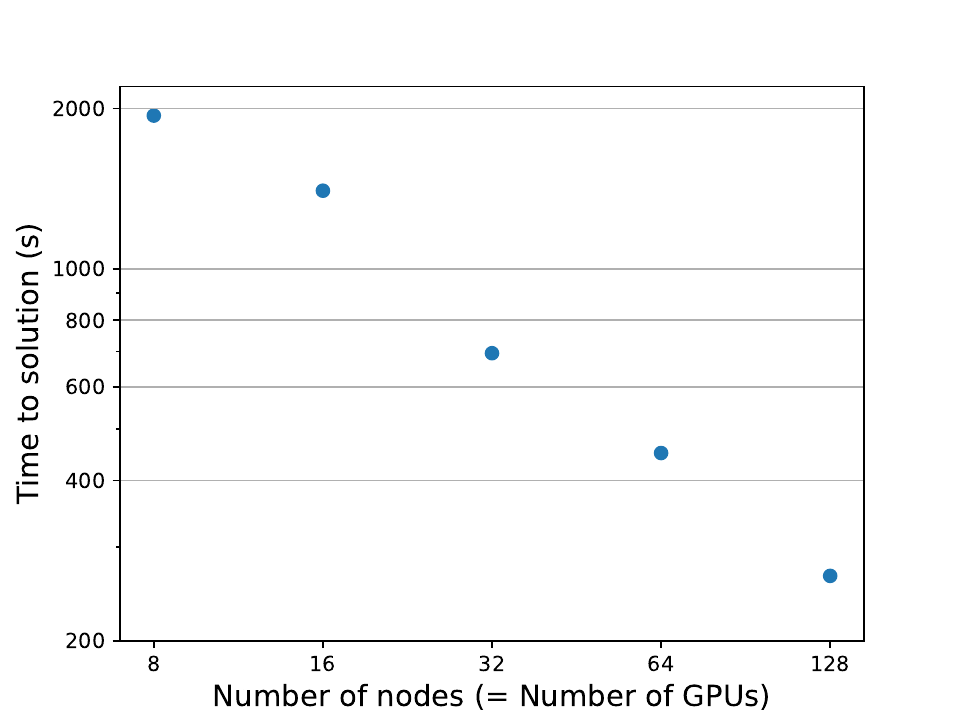}
        \caption{Time for 32-site TFIM ground state energy determination.}
        \label{fig:strong_scaling}
    \end{subfigure}
    \caption{Validation of the conjugate gradient eigensolver}
    \label{fig:eigensolver_validation}
\end{figure}

Figure~\ref{fig:ising_gap} shows the energy gap $E_1 - E_0$ as a function of $h$. Lifting of the degeneracy at $h=1$ is evident already at this system size. These calculations were performed on a single NVIDIA H100 ($96\,\mathrm{GB}$ RAM) node, thus demonstrating that the eigensolver has the expected memory footprint, and is capable of solving the ground and first excited states of nontrivial Hamiltonians (TFIM with $h \sim 1$).

Next, we verify that the eigensolver can run with distributed vectors, and investigate its time-to-solution scaling as the number of parallel nodes is increased. We consider the same TFIM Hamiltonian at $h=0.5$ with $q=32$ sites, which cannot be diagonalized on a single H100 GPU any more. The problem is then solved using 8, 16, 32, 64, and 128 nodes in parallel. Exactly identical energy values are obtained in all instances. Figure~\ref{fig:strong_scaling} plots the time to completion for computing the ground eigenpair for different parallelizations. Good strong scaling is observed.

\section{Conclusion}\label{sec:conclusion}

We presented a scalable, memory-efficient SQD solver for general Pauli Hamiltonians based on JAX. The solver consists of a general-purpose conjugate gradient eigensolver and an SQD-specific matrix-vector multiplication function that performs Hamiltonian projection internally. Both the eigensolver and the matrix-vector function support distributed vectors, allowing the overall solver to run in massively parallel GPU environments to leverage modern HPC architectures. The matrix-vector function can operate in several different caching modes, allowing control over the tradeoff between the execution time and memory footprint.

The SQD solver is currently limited to $2^{30}$ dimensions due to the need to sort the states array globally. To support the cutting-edge SQD use case with tens of billions of subspace dimensions, an upgrade to block-wise sorting is foreseen.

\begin{acks}
This work was supported by Japan Society for the Promotion of Science KAKENHI Grant Numbers JP24K07042 and JP24H00689.
\end{acks}

\section*{Code availability}

The SQD solver code is available at \url{https://github.com/UTokyo-ICEPP/rqutils/blob/main/rqutils/sqd.py} with a corresponding documentation at \url{https://rqutils.readthedocs.io/en/latest/apidoc/rqutils.sqd.html}.

\bibliographystyle{ACM-Reference-Format}
\bibliography{references}

\appendix

\section{Locally-optimal conjugate gradient}\label{app:locg}

Given a symmetric or Hermitian matrix $A$, the locally optimal conjugate gradient method for the standard minimum-eigenvalue problem
\begin{equation}\label{eq:eigenvalue_problem}
    A x = \lambda_0 x
\end{equation}
is an iterative procedure where the estimate solutions $\{x^{(0)}, x^{(1)}, \dots\}$ are formally refined through
\begin{equation}\label{eq:cg_formal_update}
    x^{(i+1)} = x^{(i)} + \alpha^{(i)} p^{(i)}
\end{equation}
with search (conjugate) directions $\{p^{(i)}\}_i$ and optimal step sizes $\{\alpha^{(i)}\}_i$. The search directions are linear combinations of the gradient of the Rayleigh quotient
\begin{equation}
    \nabla \rho(x) := \nabla \frac{x^{T} A x}{|x|^2} = 2\frac{Ax - \rho(x) x}{|x|^2}
\end{equation}
evaluated at $x \in \{x^{(i)}\}_i$, and are recursively expressed as
\begin{equation}\label{eq:search_direction_update}
    p^{(i)} = r^{(i)} + \beta^{(i)} p^{(i-1)},
\end{equation}
where
\begin{equation}
    r^{(i)} = Ax^{(i)} - \rho(x^{(i)}) x^{(i)}
\end{equation}
and
\begin{equation}
    p^{(0)} = r^{(0)}, \; p^{(-1)} = 0,
\end{equation}
using some $\{\beta^{(i)}\}_i$.

In Equation~\eqref{eq:cg_formal_update}, the step size $\alpha^{(i)}$ is optimal when $\rho(x^{(i+1)})$ is minimized. Furthermore, from Equations~\eqref{eq:cg_formal_update} and \eqref{eq:search_direction_update}, we find
\begin{equation}
    p^{(i)} \in \mathrm{span} \{x^{(i)}, x^{(i-1)}, r^{(i)}\}.
\end{equation}
Therefore, the solution update at iteration $i$ is equivalently obtained as
\begin{equation}\label{eq:cg_argmin_update}
    x^{(i + 1)} = \mathrm{arg} \min_{y \in \mathrm{span}\{x^{(i)}, x^{(i-1)}, r^{(i)}\}} \rho(y).
\end{equation}
The right-hand side of Equation~\eqref{eq:cg_argmin_update} is found by the Rayleigh-Ritz method.

Algorithm~\ref{alg:ground_locg} is the pseudocode representation of our eigensolver implementation. It includes the details on the treatment of the first two iterations, which is ommitted in the discussion above.

\begin{algorithm}
\caption{Locally optimal conjugate gradient eigensolver}\label{alg:ground_locg}
\begin{algorithmic}[1]
\Require Hermitian operator $A$ (via $x \mapsto Ax$), initial vector $x_{\mathrm{init}} \in \mathbb{C}^n$, maximum iterations $N_{\max}$, tolerance $\tau$
\Ensure Lowest eigenvalue $\theta$, eigenvector $x$, number of iterations $k$

\Function{ProjectOut}{$(b_1, \dots, b_m), v$}
  \For{$i = 0, 1, 2, 3$}
    \State $c_j \gets b_j^{\dagger} v$ for $j = 1, \dots, m$
    \State $v \gets v - \sum_{j} c_j b_j$
    \If{$i < 2$ \textbf{and} $\lVert v \rVert \neq 0$}
      \State $v \gets v / \lVert v \rVert$
    \EndIf
  \EndFor
  \If{$\lVert v \rVert < 0.99$}
    \State $v \gets 0$ \Comment{Discard numerically unreliable direction}
  \EndIf
  \State \Return $v$
\EndFunction

\Function{RayleighRitz}{$w_1, \dots, w_m$} \Comment{$m \in \{2, 3\}$}
  \State $S_{jk} \gets w_j^{\dagger} A w_k$ for $j, k = 1, \dots, m$ \Comment{$m \times m$ Hermitian}
  \State $(\theta, \kappa) \gets$ lowest eigenpair of $S$, $\lVert \kappa \rVert = 1$ \Comment{Analytic for $2\times 2$ / Cardano for $3\times 3$}
  \State \Return $(\theta, \kappa)$
\EndFunction

\Statex \textit{Iteration 0: steepest-descent residual}
\State $x \gets x_{\mathrm{init}} / \lVert x_{\mathrm{init}} \rVert$
\State $\rho \gets x^{\dagger} A x$
\State $r \gets A x - \rho\, x$

\Statex \textit{Iteration 1: two-vector Rayleigh--Ritz}
\State $p \gets r / \lVert r \rVert$ \textbf{if} $\lVert r \rVert \neq 0$ \textbf{else} $r$
\State $(\theta, \kappa) \gets \Call{RayleighRitz}{x, p}$
\State $t \gets \kappa_0 p - \kappa_1 x$
\State $u \gets \kappa_0 x + \kappa_1 p$
\State $x \gets u / \lVert u \rVert$
\State $y \gets t / \lVert t \rVert$
\State $r \gets A x - \theta x$

\Statex \textit{Main LOCG loop}
\State $k \gets 0$, \quad $\mathit{converged} \gets \textbf{false}$, \quad $\theta \gets 0$
\While{$k < N_{\max}$ \textbf{and not} $\mathit{converged}$}
  \State $p \gets \Call{ProjectOut}{(x, y), r}$
  \State $(\theta, \kappa) \gets \Call{RayleighRitz}{x, y, p}$
  \State $s \gets \kappa_1 y + \kappa_2 p$
  \State $t \gets (\kappa_0 / \lVert s \rVert)\, s - \lVert s \rVert\, x$ \Comment{Orthogonal to new $x$, spans old $\{x, s\}$}
  \State $u \gets \kappa_0 x + s$
  \State $x \gets u / \lVert u \rVert$
  \State $y \gets t / \lVert t \rVert$
  \State $v \gets A x$
  \State $r \gets v - \theta x$
  \State $\delta \gets 10\, n\, (\lVert v \rVert - \theta)$ \Comment{Relative tolerance scale}
  \State $\mathit{converged} \gets \lVert r \rVert < \tau \delta$
  \State $k \gets k + 1$
\EndWhile
\State \Return $(\theta, x, k)$
\end{algorithmic}
\end{algorithm}

\end{document}